# Tunable nonlinear optical bistability based on Dirac semimetal in photonic crystal Fabry-Perot cavity

Huayue Zhang[1,2], Xueyan Tang[2], Hongxia Yuan[2], Xin Long[2], Haishan Tian[2], Xinghua Wu[3], Zhijun Guo[1,*], and Leyong Jiang[2,†]

[1]Institute of Information Science and Technology, Hunan Normal University, Changsha 410081, China;

[2]School of Physics and Electronics, Hunan Normal University, Changsha 410081, China;

[3]Key Laboratory for Microstructural Functional Materials of Jiangxi Province, College of Science, Jiujiang University, Jiujiang 332005, China.

Corresponding Author: *guozhijun@hunnu.edu.cn and †jiangly28@hunnu.edu.cn

## Abstract

In this paper, we study the nonlinear optical bistability (OB) in a symmetrical multilayer structure. This structure is constructed by embedding a nonlinear three-dimensional Dirac semimetal (3D DSM) into a solution filled one-dimensional photonic crystal Fabry-Perot cavity. OB stems from the third order nonlinear conductivity of 3D DSM and the local field of resonance mode could enhance the nonlinearity and reduce the thresholds of OB. This structure achieves the tunability of OB due to that the transmittance could be modulated by the Fermi energy. OB threshold and threshold width could be remarkably reduced by increasing the Fermi energy. Besides, it is found that the OB curve depends heavily on the angle of incidence of the incoming light, the structural parameters of the Fabry-Perot cavity, and the position of 3D DSM inside the cavity. After parameter optimization, we obtained OB with a threshold of $10^6$ V/m. We believe this simple structure provides a reference idea for realizing low threshold and tunable all optical switching devices.

# 1. Introduction

Optical bistability (OB) is a typical nonlinear optical phenomenon, which refers to the nonlinear phenomenon where an input light intensity corresponds to two output light intensities that can be converted into each other within a certain range, and at this time, there is a hysteresis loop similar to the "S" shape between the input and output light intensities [1,2]. Due to the nonlinear characteristics similar to the hysteresis curve, OB has been widely used in fields such as optical communication and photonic computing, such as biosensors [3], all-optical switching [4], optical transistors [5], and optical memories [6]. Currently, the main goal of studying OB is to reduce the threshold of OB to achieve practical OB devices. Generally, researchers' methods to reduce OB threshold can be classified as two categories in rough: the first category is to take advantage of resonant structures such as photonic crystal micro-cavities [7,8], Fabry-Perot (FP) cavities [9,10], fiber bragg grating cavities [11,12], and surface plasmon resonance [13-15] to lower the OB threshold by enhancing the localized field enhancement effect; the second category is to use nonlinear materials with high nonlinear coefficients and add them to micro-nano structures to enhance the nonlinear effect of the system and achieve the goal of reducing OB threshold. For example, micro-ring resonators based on cubic silicon carbide and photonic circuits based on sulfide glass have been reported [16,17]. In addition, studies have shown that graphene has superior optoelectronic properties and can fully utilize its advantages in the field of OB, with its most prominent feature being a huge nonlinear coefficient. Furthermore, graphene has characteristics such as tunable conductivity and ultra-fast modulation speed, making it a favorite of researchers [18,19]. Currently, researchers are eagerly reporting on the method of reducing the threshold of OB by adding graphene to composite structures [20-22]. Recently, people have turned their attention to another type of topological material based on graphene, the three-dimensional Dirac semimetal (3D DSM). 3D DSM is a bulk medium and its volume electrons form a three-dimensional Dirac cone, which is also known as "three-dimensional graphene". Although 3D DSM is called "three-dimensional graphene", unlike the two-dimensional Dirac fermions on the surface of graphene, 3D DSM has three-dimensional Dirac fermions in the volume. The unique electronic structure of 3D DSM gives it many peculiar properties. Previous reports have shown that 3D DSM not only has similar properties to graphene in controlling the dielectric constant by adjusting the Fermi energy but also exhibits high electron mobility [23], good stability, strong coupling with light, and the ability to avoid the huge metal losses that occur in traditional metal metamaterials [24]. It is worth noting that the nonlinear refractive index coefficient of 3D DSM in the terahertz range is relatively high [25], so significant OB can be observed at lower incident electric field ranges. Based on these advantages, 3D DSM material stands out as a candidate material for constructing nonlinear optical devices in the field of OB, providing a new perspective for designing flexible and controllable nonlinear OB devices.

As is well known, the simplest device for implementing OB is the FP cavity structure, which can create the conditions necessary for the occurrence of OB by filling the cavity with a nonlinear medium. By increasing the intensity of the incident light, the refractive index of the nonlinear medium in the FP cavity changes, resulting in a bistable relationship between input and output light intensities. Traditionally, the FP cavity consists of two mirrors, but with the development of one-dimensional photonic crystal (1D PC), researchers have found that the photonic band gaps and localization properties of 1D PC have potential applications in the field of OB. Therefore, the use of two 1D PCs instead of two mirrors in the traditional FP cavity to achieve OB has attracted widespread attention. Currently, many optical devices based on photonic crystal FP cavity structures have been reported, such as optical absorption, optical filters, and nonlinear effects. Therefore, we imagine what changes would occur in the cavity if we use two 1D PCs instead of the two mirrors in the FP cavity and fill the cavity with 3DDSM. Will hysteresis phenomenon still occur in such a new FP cavity? And what impact will 3D DSM have on hysteresis phenomenon? Based on this, we theoretically propose a photonic crystal FP cavity based on 3D DSM, which is placed at the center of the FP cavity. We found that 3D DSM can provide strong nonlinear effects for the structure, and the unique band structure of 1D PC can increase the flexibility of controlling OB with effect. Moreover, the threshold and hysteresis width of OB can be regulated by tuning parameters such as the Fermi energy of 3D DSM, incident angle, relaxation time, period of 1D PC, length of FP cavity, and position of 3D DSM in the cavity. Therefore, we believe that the photonic crystal FP cavity structure based on 3D DSM has broad potential applications in the field of nonlinear terahertz devices and nanophotonics, providing a reference for achieving low-threshold and tunable OB.

## 2. Theoretical Model and Method

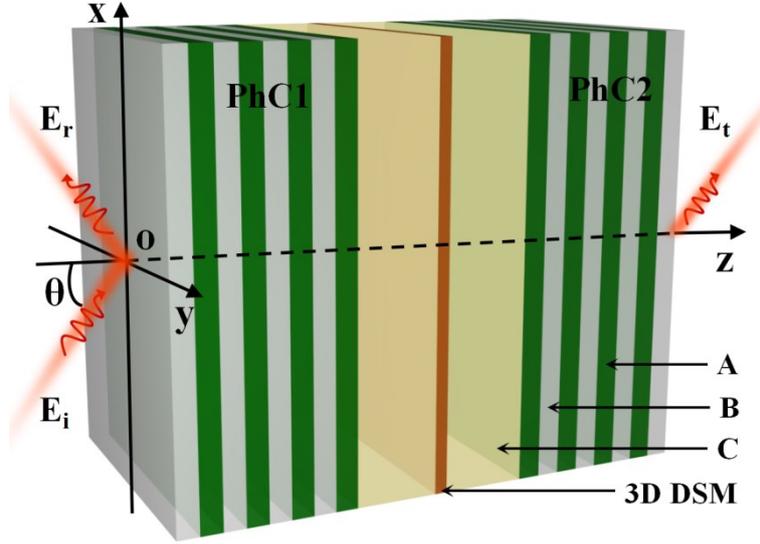

Fig.1 Schematic diagram of a photonic crystal FP cavity based on a 3D DSM.

It is considered to use a FP cavity composed of two symmetric photonic crystals, filled with 3D DSM and water solution. As shown in Fig. 1, the 3D DSM is placed in the middle of the cavity, and the medium C placed on both sides of the 3D DSM is a water solution with a refractive index of $n_c = 1.33$ and a thickness of $d_c$. The photonic crystal is composed of dielectric materials A and B with a cycle of $m = 4$, where the refractive index of dielectric material A is $n_a = 1.5$, and dielectric material B is glass with a refractive index of $n_b = 2.3$ [29,30]. The thicknesses of A and B are $d_a$ and $d_b$, respectively. The central wavelength is set to $\lambda_c = 300\,\mu\text{m}$, and the length of the FP cavity is set to $L = 150\,\mu\text{m}$. The thickness of dielectric material A and B is set to $d_{a,b} = \lambda_c/4n_{a,b}$, and the thickness of medium C is also set to $d_c = \lambda_c/4n_c$. The initial parameters of 3D DSM are set to $E_F = 1\,\text{eV}$, $\tau = 0.2\,\text{ps}$, and $d_D = 40\,\text{nm}$. The direction perpendicular to the photonic crystal is set as the z direction, and the incident electromagnetic wave is assumed to be TE polarization. It should be noted that the technology for preparing such multilayer dielectric structures is relatively mature, and because the 3D DSM is a bulk medium, the technical difficulty in manufacturing is reduced. Overall, it is feasible to apply the structure proposed in this paper in practice.

Ignoring the effects of external magnetic fields and random phase approximation, we can use the semiclassical Boltzmann transport equation with relaxation time approximation to calculate the 3D DSM linear intraband optical conductivity in the terahertz range, as expressed in equation (1) [31]:

$$\sigma^{(1)} = \sigma_0 \frac{4}{3\pi^2} \frac{\tau}{1-i\omega\tau} \frac{(k_B T)^2}{\hbar^2 v_F} \left[ 2Li_2\left(-e^{-\frac{E_F}{k_B T}}\right) + \left(\frac{E_F}{k_B T}\right)^2 + \frac{\pi^2}{3} \right], \quad (1)$$

here $\sigma_0 = e^2/4\hbar$, $\tau$ is the relaxation time of the 3D DSM, $k_B$ is the Boltzmann constant, $T$ is temperature, $\hbar$ is the reduced Planck constant, $v_F = 10^6$ m/s is the Fermi velocity of the electrons, $Li_s(z)$ is the polylogarithm function, and $E_F$ is the Fermi energy of the 3D DSM. The third-order nonlinear optical conductivity of the 3D DSM is expressed as equation (2) [31]:

$$\sigma^{(3)} = \sigma_0 \frac{8e^2 v_F}{5\pi^2 \hbar^2} \frac{\tau^3}{(1+\omega^2\tau^2)(1-2i\omega\tau)} \frac{1}{1+\exp(-E_F/k_B T)}, \quad (2)$$

therefore, we can write the third-order polarization of the 3D DSM as the following expression:

$$\chi^{(3)} = i\sigma^{(3)}/\varepsilon_0 \omega, \quad (3)$$

here, $\varepsilon_0$ represents the permittivity of vacuum. In addition, The linear complex refractive index of the 3D DSM is denoted as:

$$n_D = n + ik = \sqrt{1 + i\sigma^{(1)} / \varepsilon_0 \omega}, \tag{4}$$

from the above expressions, it can be found that the third-order polarizability $\chi^{(3)}$ of 3D DSM is bound up with the linear in-band optical conductivity $\sigma^{(1)}$, and the linear complex refractive index $n_D$ is bound up with the third-order nonlinear optical conductivity $\sigma^{(3)}$. while regulating the Fermi energy $E_F$ and relaxation time $\tau$ can regulate $\sigma^{(1)}$ and $\sigma^{(3)}$. This provides convenient circumstance for effectuating a tunable low-threshold OB.

The reflection-transmission relationship of the entire construction is estimated making use of the transfer matrix method. As shown in Fig. 1, we set the light beam to propagate along the z-axis, with the 3D DSM parallel to the x-axis, and the 3D DSM placed at $z = 0$. The transmission matrix of the photonic crystal and the transmission matrix of the aqueous tier inside the cavity can be expressed as:

$$M_j = \begin{bmatrix} \cos(k_{jz}d_j) & -i\frac{k_0}{k_{jz}}\sin(k_{jz}d_j) \\ -i\frac{k_{jz}}{k_0}\sin(k_{jz}d_j) & \cos(k_{jz}d_j) \end{bmatrix}, \tag{5}$$

here $j = \{A, B, S\}$, $k_{jz} = \sqrt{(k_0 n_j)^2 - (k_y)^2}$ is the constituent of medium j disseminating follow the z-axis, $k_0 = \omega/c$ is the wave vector in vacuum, $d_j$ is the thickness of medium j, $k_y = k_0 \sin\theta$ is the constituent of the incident wave disseminating follow the y-axis, and $\theta$ is the incident angle.

And then, we canvass the transmission matrix of the nonlinear medium layer, which is the 3D DSM layer in this article. The transmission matrix of a Kerr-type nonlinear material can be expressed as the following equation [32]:

$$M_D = \frac{k_0}{k_{z+} + k_{z-}} \begin{bmatrix} \frac{k_{z-}}{k_0}\exp(-ik_{z+}d_D) + \frac{k_{z+}}{k_0}\exp(ik_{z-}d_D) & \exp(-ik_{z+}d_D) - \exp(ik_{z-}d_D) \\ \frac{k_{z-}k_{z+}}{k_0^2}\left[\exp(-ik_{z+}d_D) - \exp(ik_{z-}d_D)\right] & \frac{k_{z+}}{k_0}\exp(-ik_{z+}d_D) + \frac{k_{z-}}{k_0}\exp(ik_{z-}d_D) \end{bmatrix}, \quad (6)$$

here, $k_{z+}$ is the z-component of the forward-propagating wave vector, $k_{z-}$ is the z-component of the backward-propagating wave vector, and their expressions are given by equation (7):

$$k_{z\pm} = \sqrt{(k_0 n_D)^2 - (k_y)^2}\left(1 + U_\pm + 2U_\mp\right)^{\frac{1}{2}}, \quad (7)$$

with $U_\pm = \frac{k_0^2 \chi^{(3)}}{n_D^2 k_0^2 - k_y^2}|A_\pm|^2$. Where $A_+$ and $A_-$ are the amplitude of vibration of the forward and backward waves, separately. The wave vectors $k_{z+}$ and $k_{z-}$ can be obtained by disposing of the coupled nonlinear equations below with the aid of the following equation:

$$\begin{pmatrix} U_+ \\ U_- \end{pmatrix} = \left|\begin{pmatrix} 1 & 1 \\ \frac{k_{z+}}{(k_0\mu_D)} & \frac{-k_{z-}}{(k_0\mu_D)} \end{pmatrix}^{-1} (M_B M_A)^m \begin{pmatrix} 1 \\ p \end{pmatrix}\right|^2 U_t, \quad (8)$$

where $U_t$ is the transmitted intensity, $p = \sqrt{\varepsilon_0\mu_0 k_0^2 - \beta^2}/(k_0\mu_0)$. If we can calculate a set of coupled nonlinear equations about $U_+$ and $U_-$ through fixed-point iteration, we can obtain the specific expression of the transmission matrix $M_D$ of the 3D DSM layer [33]. First, we set the initial value $U_\pm = 0$, and then we seek the stable solutions of $U_+$ and $U_-$ through multiple iterations. Once the solution of $U_\pm$ is determined, we can calculate the transmission matrix $M_D$ of the 3D DSM layer using Eq. (6) and Eq. (7). Therefore, the transmission matrix of the entire structure can be determined as:

$$M = (M_B \times M_A)^m \times M_C \times M_D \times M_C \times (M_A \times M_B)^m, \quad (9)$$

the transmission coefficient of the structure can be worked out as follow:

$$t = \frac{2p_f}{\left[M_{11} + M_{12}p_f\right]p_f + \left[M_{21} + M_{22}p_f\right]}, \tag{10}$$

here $p_f = \left(k_0^2 - k_y^2\right)^{\frac{1}{2}}/k_0$, $M_{ij}$ is an element of the $2\times 2$ matrix $M$. Finally, we can obtain the relevance between $E_t$ and $E_i$, and observe the obvious OB phenomenon by adjusting the parameters.

## 3. Results and Discussions

In this section, we first talk over the relevance between the transmitted electric field $E_t$ and the incident electric field $E_i$ under various Fermi energies in 3D DSM, while keeping the other values keep up with those mentioned in Fig. 1. The key factor in realizing OB in this study is the large nonlinear conductivity of Dirac semimetals. Specifically, when we fill 3D DSM into the FP cavity, we observe a significant change in the relevance curve between $E_t$ and $E_i$, resulting in a hysteresis loop in the transmission curve. This is because the conditions required to achieve OB become easier to satisfy when the 3D DSM are filled into the FP cavity. From Fig. 2(a), we can see that when $E_i$ is small, the transmitted electric field $E_t$ gradually raises with the incident electric field $E_i$ until a specific value, at which point the transmitted electric field $E_t$ undergoes a jump. However, when $E_i$ is large, $E_t$ gradually decreases with $E_i$ up to a specific value, at which point $E_t$ undergoes a jump. In the case of the double stable curve at $E_F$=0.8eV in the figure, when $E_i$ gradually increases from a small value to $|E_i| = 2\times 10^7$ V/m, the transmitted electric field $E_t$ undergoes an upward mutation, and the value of $|E_i| = 1\times 10^7$ V/m at this point is called the upper threshold field of the switch. However, when the incident electric field $E_i$ gradually decreases from a large value to $|E_i| = 1\times 10^7$ V/m, the transmitted

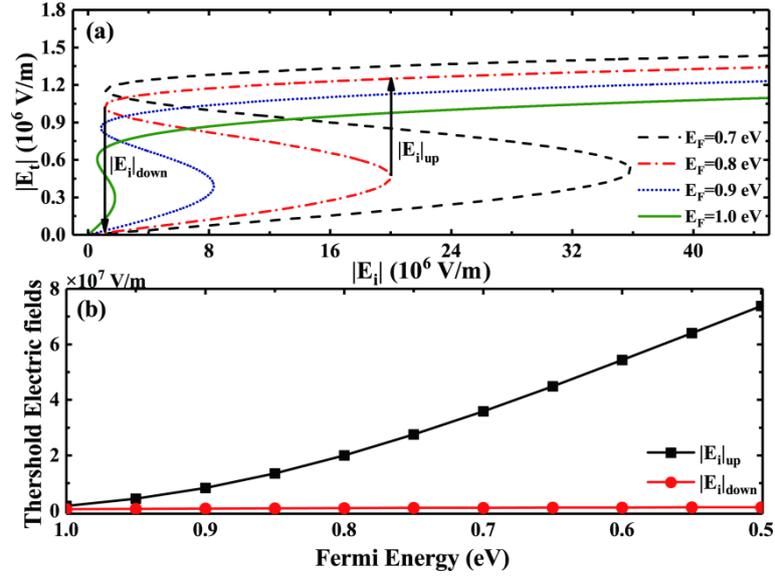

Fig.2 (a) Dependence of $|E_t|$ on $|E_i|$ for different Fermi energy $E_F$; (b) influence of the Fermi energy $E_F$ of the 3D DSM on the electric field threshold.

electric field $E_t$ undergoes a downward mutation, and the value of $|E_i|_{down} = 1 \times 10^7 \, \text{V/m}$ at this point is called the lower threshold field of the switch. In addition, the hysteresis width is denoted as $\Delta |E_i| = |E_i|_{up} - |E_i|_{down} = 1 \times 10^7 \, \text{V/m}$. Moreover, from Fig. 2(a), we can also observe that as the Fermi energy $E_F$ of the 3D DSM increases, both the upper and lower threshold fields of the OB ($|E_i|_{up}$ and $|E_i|_{down}$) raise, and they move to higher light intensity ranges to the right. We describe this phenomenon in detail in Fig. 2(b). As shown in the figure, as $E_F$ of the 3D DSM increases, the hysteresis loop between $E_t$ and $E_i$ gradually increases, and its hysteresis width also significantly widens because the speed of the upper threshold field of the OB moving to the right is much faster than that of the lower threshold field of the OB. Additionally, we can find that although reducing $E_F$ of the 3D DSM is advantageous for lowering the OB threshold, when the Fermi energy $E_F$ decreases to a certain extent, the hysteresis width disappears, and the OB curve vanishes as well. In summary, the discovery of the above laws is beneficial for us to regulate the OB threshold and hysteresis width by adjusting $E_F$ of the 3D DSM, providing a pathway for people to regulate the OB phenomenon through external electric field control.

Next, we will discuss the impact of incident angle variations on the OB curve. We set the Fermi energy as $E_F = 1\,\text{eV}$, and other values keep up with Fig. 1. As illustrated in Fig. 3, the threshold size and hysteresis width of the OB change with the incident angle, indicating that the OB phenomenon is highly sensitive to the incident angle, which follows a similar pattern to the discussion on the 3D DSM Fermi energy $E_F$ in the previous section. As shown in Fig. 3(a), increasing the incident angle of the structure causes the transmittance curve to shift towards higher light intensities, and the hysteresis width gradually increases with increasing incident angle. Meanwhile, the dependence of $E_t$ on $E_i$ also exhibits a similar variation tendency, the above rule is reflected in Fig. 3(b). We take $\theta=5°$ and $\theta=10°$ as examples to analyze the changing pattern in detail: when $\theta=5°$, $|E_i|_{up} = 2.98 \times 10^6\,\text{V/m}$, $|E_i|_{down} = 6.60 \times 10^5\,\text{V/m}$, the hysteresis width $\Delta|E_i| = 2.32 \times 10^6\,\text{V/m}$; when $\theta = 10°$, $|E_i|_{up} = 4.34 \times 10^6\,\text{V/m}$, $|E_i|_{down} = 8.05 \times 10^6\,\text{V/m}$, the hysteresis width is $\Delta|E_i| = 3.54 \times 10^5\,\text{V/m}$. These calculated results confirm the above pattern well. In addition, as shown in the OB line at $\theta = 0°$ in Fig. 3(b), although the hysteresis width gradually narrows with decreasing incident angle, even when the incident angle decreases to $0°$, there is still a hysteresis width in the relevance curve between $E_t$ and $E_i$. Therefore, it can be proved that the method of using the 3D DSM-based photonic crystal FP cavity to achieve low-threshold and tunable OB can also be implemented under vertical incidence conditions without requiring specific incident angles. Compared to certain traditional methods of achieving low-threshold OB, this method has more advantages and is easier to implement.

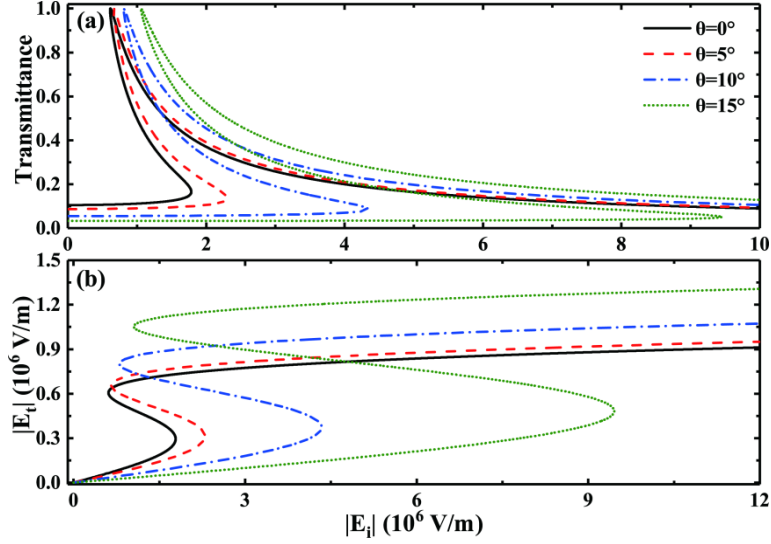

Fig.3 Under various incidence angles, the relationship between (a) transmittance, (b) $|E_t|$ and $|E_i|$.

Furthermore, we investigate the influence of relaxation time and photonic crystal period variations on the OB phenomenon while keeping other relevant values remain unchanged. On the one hand, we observe that the change in the relevance between $E_t$ and $E_i$ exhibits the same trends as those seen in the OB threshold and hysteresis width for different relaxation times. In the wake of $\tau$ decreases, the OB threshold also decreases. However, as $|E_i|_{up}$ and $|E_i|_{down}$ both gradually increase, but $|E_i|_{down}$ increases faster than $|E_i|_{up}$, the hysteresis width rapidly narrows. For instance, when $\tau = 0.4 \, \text{ps}$, $|E_i|_{up} = 4.02 \times 10^7 \, \text{V/m}$, $|E_i|_{down} = 1 \times 10^6 \, \text{V/m}$, the hysteresis width is $\Delta|E_i| = 3.92 \times 10^7 \, \text{V/m}$, and when $\tau = 0.2 \, \text{ps}$, $|E_i|_{up} = 1.78 \times 10^6 \, \text{V/m}$, $|E_i|_{down} = 6.11 \times 10^5 \, \text{V/m}$, it is $\Delta|E_i| = 1.17 \times 10^6 \, \text{V/m}$. Simulation results reveal that the OB curve vanishes when the relaxation time expends to $\tau=0.1\,\text{ps}$. Therefore, while reducing the relaxation time can effectively lower the OB threshold, it cannot be reduced to an excessively low level, which will cause the OB phenomenon to disappear. On the other hand, we find that the impact of photonic crystal period variations on the OB phenomenon is highly similar to that of relaxation time. The OB threshold decreases as the photonic crystal period decreases, and the hysteresis width between $E_t$ and $E_i$ narrows. Thus, careful consideration is required in choosing both the relaxation time and photonic crystal period to ensure optimal OB phenomenon production and the fabrication of more reasonable and practical OB devices.

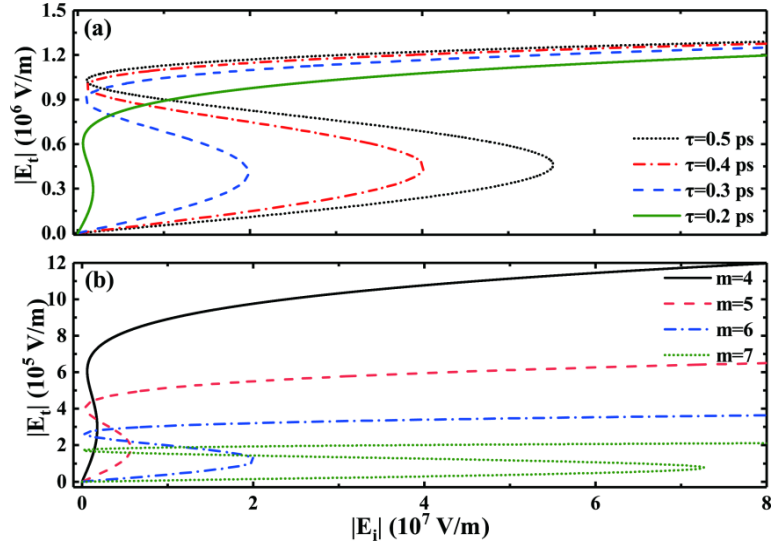

Fig.4 The relationship between $E_t$ and $E_i$ at (a) different relaxation times and (b) different photonic crystal periods.

Regarding the relevance between $E_t$ and $E_i$, we observed that the OB phenomenon is not only sensitive to the relaxation time, photonic crystal period, and 3D DSM Fermi energy, but also to the cavity length of the FP cavity and the position of the 3D DSM layer in the cavity. Firstly, with the cavity length fixed at $L=150\,\mu\text{m}$,

we studied the effect of the placement of the 3D DSM layer in the cavity on the OB

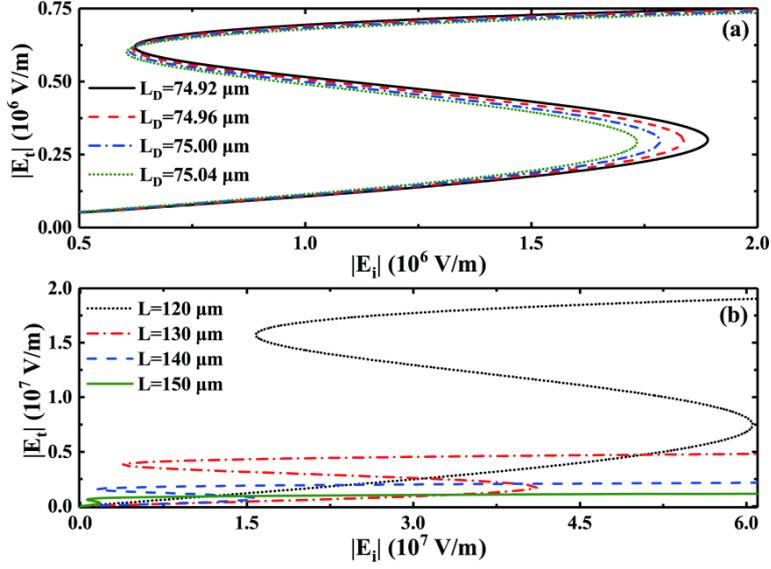

Fig.5 (a) Variation of the cavity length and (b) relevance between $E_t$ and $E_i$ for different positions of the 3D DSM layer inside the cavity.

phenomenon. The position of the center point of the 3D DSM layer can be represented as $L_D=75\,\mu m$. As shown in Fig. 5(a), as the position of the 3D DSM layer in the FP cavity changes, the resulting OB phenomenon also changes, with a corresponding change in the threshold size and hysteresis width, thus achieving flexible control of the OB phenomenon by adjusting the position of the 3D DSM layer. Additionally, the cavity length of the FP cavity should satisfy $kL=\pi$, where the frequency is inversely proportional to the cavity length $L$. Therefore, by adjusting the cavity length $L$, the OB phenomenon can be controlled. Fig. 5(b) shows the changes in the hysteresis curve after changing the cavity length. It can be observed that as the cavity length increases, the threshold size and hysteresis width of the OB phenomenon both increase significantly. Thus, we believe that the cavity length of the FP cavity can be an important means of controlling the OB phenomenon.

## 4. Conclusions

In conclusion, we have studied the OB phenomenon in a photonic crystal FP cavity based on 3D DSM. By utilizing the unique band structure of the photonic crystal, the resonant effect of the FP cavity, and the extraordinary nonlinear properties of 3D DSM, we have achieved low-threshold OB at terahertz frequencies, which can be flexibly controlled in terms of threshold size and hysteresis width. Our theoretical studies have shown that embedding 3D DSM in a FP cavity can significantly enhance the hysteresis behavior. By adjusting parameters such as the Fermi energy of the 3D DSM, the incident angle of the beam, the relaxation time, the cavity length of the FP cavity, and the position of the 3D DSM layer in the FP cavity, the OB phenomenon can be further controlled. Through numerical simulations, we have found suitable 3D DSM and structural parameters and obtained an OB curve with a threshold of $10^6$ V/m. Our research provides new ideas for achieving tunable low-threshold OB at terahertz frequencies. We are firmly convinced that the construction came up with in this article for realizing OB is relatively simple and has advantages in the manufacture of practical OB devices, and has an extensive range of applications in the field of nonlinear optical devices.

## Acknowledgments

This work was supported by the Scientific Research Fund of Hunan Provincial Education Department (Grant No. 21B0048), Natural Science Foundation of Hunan Province (Grant Nos.2022JJ30394), the Changsha Natural Science Foundation (Grant Nos. kq2202236, kq2202246), and the Science and Technology Project of Jiangxi

Provincial Education Department (Grant No. GJJ190911).